\documentclass[12pt,a4paper]{article}

\catcode`\@=11
\@addtoreset{equation}{section}

\global\arraycolsep=2pt
\usepackage{mathrsfs,amsbsy,amssymb,latexsym,amsfonts,amsmath}
\usepackage{graphicx,color,cite}

\allowdisplaybreaks

\newcommand{\dd}{{\rm d}}

\newcommand{\gs}{g_{\text{s}}}

\newcommand{\tf}{T_{\rm F} }

\newcommand{\rc}{r_{\rm c} }

\newcommand{\rt}{r_{\rm t} }

\usepackage[hypertex]{hyperref}

\begin{document}

\begin{flushright}
\parbox{4.2cm}
{KUNS-2453}
\end{flushright}

\vspace*{2cm}

\begin{center}
{\Large \bf Holographic Schwinger effect in confining phase}
\vspace*{2cm}\\
{\large Yoshiki Sato\footnote{E-mail:~yoshiki@gauge.scphys.kyoto-u.ac.jp} 
and 
Kentaroh Yoshida\footnote{E-mail:~kyoshida@gauge.scphys.kyoto-u.ac.jp} 
}
\end{center}

\vspace*{1cm}
\begin{center}
{\it Department of Physics, Kyoto University \\ 
Kyoto 606-8502, Japan} 
\end{center}

\vspace{1cm}

\begin{abstract}
We consider the Schwinger effect in confining phase by using a holographic setup. 
The potential analysis is performed for the confining D3-brane and 
D4-brane backgrounds. We find the critical electric field above which 
there is no potential barrier and the system becomes unstable catastrophically. 
An intriguing point is that no Schwinger effect occurs when the electric field 
is smaller than the confining string tension. 
\end{abstract}

\thispagestyle{empty}
\setcounter{page}{0}

\newpage

\section{Introduction}

In the vacuum of quantum electromagnetic dynamics (QED), 
virtual electron and positron pairs are created and annihilated, momentarily and continuously.  
The pairs can be materialized in a strong electric field. 
This is a non-perturbative phenomenon and is known as the Schwinger effect \cite{Schwinger} 
(For related progress see \cite{AAM,AM}). 
This is not intrinsic to the original QED but ubiquitous in quantum field theories coupled to a $U(1)$ gauge field. 
In all cases, we refer it as to the Schwinger effect for convenience sake. 

\medskip 

It is interesting to consider the Schwinger effect in the context of 
the AdS/CFT correspondence \cite{M,GKP,W}. First, we have to realize 
a system coupled with a $U(1)$ gauge field.
It can be realized basically via the Higgs mechanism with the breaking of the gauge group 
from $SU(N+1)$ to $SU(N)\times U(1)$\,. Then the fundamental scalar fields, 
which belong to the W-boson supermultiplet and are often called ``W-bosons'' or ``quarks'', 
are coupled to a $U(1)$ gauge field as well as an $SU(N)$ one (We will call them W-bosons hereafter). 
The Coulomb potential between them 
is also computed in the holographic setup \cite{Wilson1,Wilson2}.
Hence we can consider the pair production rate of the W-bosons via the Schwinger effect \cite{GSS,SZ}. 

\medskip 

In the setup proposed by Semenoff and Zarembo \cite{SZ},  
the probe D3-brane is located far from the boundary so as to make 
the mass of W-bosons lighter, rather than infinitely heavy.  
The exponential factor in the production rate of W-bosons is evaluated 
from the string world-sheet attaching to the probe D3-brane. 
The string solution is obtained by terminating the one for a circular Wilson loop \cite{BCFM,DGO} 
at the location of the probe D3-brane. Then the value of a critical electric field,  
for which the potential barrier vanishes, agrees with the one obtained from the Dirac-Born-Infeld (DBI) action. 
This is compatible with the stringy Schwinger effect \cite{max1,max2}. The setup in \cite{SZ} 
has been generalized to the pair production of monopole-antimonopole pairs and dyon-antidyon pairs 
\cite{BKR} and the case with magnetic fields \cite{SY}. 

\medskip 

The same result on the critical electric field has also been reproduced from the potential analysis \cite{SY2}. 
The advantage of this procedure is that we do not have to take account of fluctuations 
around a circular Wilson loop (For attempts to evaluate the fluctuations around a circular Wilson loop, 
see \cite{DGT,AmMa,KM}). Thus the results in \cite{SY2} give a strong support for the 
proposal of Semenoff and Zarembo from another perspective. 

\medskip 

In this letter we will apply the potential analysis to confining theories with a holographic setup. 
Although there are various confining backgrounds, we will focus upon here 1) D3-brane and 
2) D4-brane backgrounds, where one of spatial directions is compactified on an S$^1$ circle 
with the (anti-)periodic boundary conditions for bosons (fermions) \cite{Witten}. 
For both cases, the total potential has two critical values 
of the electric field. The first is the same as the one in the Coulomb phase.  
The potential barrier vanishes at the critical value and the vacuum becomes unstable catastrophically. 
The second one is intrinsic to the confining phase. The critical value of the electric field 
agrees with the confining string tension. Below this value, the potential tends to diverge 
at infinitely long distances and thus no Schwinger effect occurs. 
When the electric field dominates the confining string tension, the potential vanishes at infinity 
and hence the Schwinger effect can occur as a tunneling process.

\section{Confining D3-brane background}

Let us perform the potential analysis for the confining D3-brane background. 

\medskip 

The background metric in the Lorentzian signature is given by \cite{HM} 
\begin{align}
\dd s^2&=g_{\mu\nu}\dd x^{\mu}\dd x^{\nu}  \notag \\ 
& =\frac{r^2}{L^2} \left( -(\dd x^0)^2+\sum _{i=1}^2(\dd x^i )^2+f(r)(\dd x^3 )^2\right)
+\frac{L^2}{r^2}f(r)^{-1}\, \dd r^2+L^2 \dd \Omega _5 ^2\,, \label{metric}
\end{align}
where the ten-dimensional spacetime coordinates are described by $x^{\mu}~(\mu=0,\ldots,9)$
and $\dd\Omega_5$ is the line element of S$^5$ with the unit radius. 
The AdS radius $L$ is related to the gauge-theory parameter 
as $L^2 = \sqrt{\lambda}\,\alpha'$ ($\lambda \equiv 4\pi \gs N$:~'t~Hooft coupling). 
Note that the metric (\ref{metric}) contains the scalar function 
\begin{eqnarray}
f(r) \equiv 1-\frac{\rt^4}{r^4}\,. 
\end{eqnarray}
Here $\rt~(\geq 0)$ is interpreted as the inverse compactification radius in the $x^3$-direction. 
As $\rt$ grows, the radius tends to shrink. Also, $\rt$ plays the role like ``temperature''. 
When $\rt =0$ (zero temperature)\,, the usual AdS$_5\times {\rm S}^5$ background is reproduced.

\medskip 

We are concerned with the classical solution of the fundamental string on the background (\ref{metric}). 
We will work in the Euclidean signature after performing a Wick rotation to the metric (\ref{metric}). 
The Nambu-Goto (NG) string action is given by 
\begin{align}
S &= \tf \int \!\! \dd \tau \!\! \int\!\!  \dd \sigma \, \mathcal{L} 
\nonumber \\ &
= \tf \int \!\! \dd \tau \!\! \int\!\!  \dd \sigma \, \sqrt{\det G_{ab}}\,, \qquad  G_{ab} \equiv  
\frac{\partial x^{\mu}}{\partial \sigma ^a}\frac{\partial x^{\nu}}{\partial \sigma ^b}\, g_{\mu \nu}\,. 
\end{align}
Here $\tf=1/2\pi\alpha'$ is the string tension and the string world-sheet coordinates are described by $\tau$ and $\sigma$\,. 

\medskip 

\begin{figure}[htbp]
\begin{center}
\includegraphics[scale=.5]{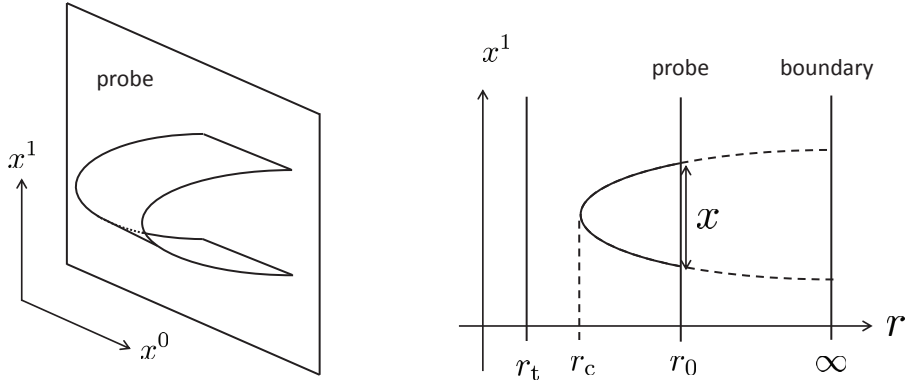}
\caption{\footnotesize The configuration of the string world-sheet. \label{AdS-soliton:fig}}
\end{center}
\end{figure}

We work with the static gauge 
\[
x^0=\tau \,,\qquad x^1=\sigma 
\]
and suppose the ansatz for the radial direction,  
\[
r=r(\sigma)\,, \qquad \mbox{the others are constant.}
\]
Then the ansatz leads to the following expression,  
\begin{equation}
{\cal L} =\sqrt{\frac{1}{1-\rt^4/r^4}\left(\frac{\dd r}{\dd \sigma}\right)^2+\frac{r^4}{L^4}}\,. 
\label{L}
\end{equation}
The analysis below is almost parallel to the one in \cite{SY2}. 
Now that ${\cal L} $ does not depend on $\sigma$ explicitly, 
one can obtain the conserved quantity,
\begin{eqnarray}
\frac{\partial \mathcal{L}}{\partial (\partial_{\sigma}r)}\partial_{\sigma}r  - \mathcal{L}\,.
\end{eqnarray}
By imposing the boundary condition at $\sigma=0$\,,  
\begin{equation}
\frac{\dd r}{\dd \sigma} =0\,, \qquad  r=\rc~~~~(\rt < \rc < r_0)\,,  
\label{bc}
\end{equation}
the conserved quantity is written as 
\begin{align}
\frac{r^4}{\displaystyle \sqrt{\frac{1}{1-\rt^4/r^4}\left(\frac{\dd r}{\dd \sigma}\right)^2+\frac{r^4}{L^4}} }= \mbox{constant} \equiv 
\rc ^2L^2\,, \label{conserved}
\end{align}
Equation \eqref{conserved} can be rewritten as 
\begin{align}
\frac{\dd r}{\dd \sigma}=\frac{1}{L^2}\sqrt{(r^4-\rt^4)\left(\frac{r^4}{\rc ^4}-1\right)}\,. \label{diff}
\end{align}
The configuration we consider is depicted in Fig.\,\ref{AdS-soliton:fig}. 
It would be worth noting the relation between the present configuration and the previous one discussed 
in \cite{SY2}. In the previous analysis \cite{SY2}, a temporal Wilson loop is considered with 
the (Euclidean) black three-brane background \cite{HS}, 
while now a temporal Wilson loop is discussed in the confining background (\ref{metric}). 
When we consider a spatial Wilson loop in the (Euclidean) black three-brane background, 
the analysis completely agrees with the present one. 

\medskip 

By solving the differential equation (\ref{diff}) under the boundary condition (\ref{bc}),  
the distance $x$ between W-boson and anti W-boson is obtained as 
\begin{align}
x
&=\frac{2L^2}{r_0a}\int_1^{1/a} \! \! \frac{\dd y}{\sqrt{(y^4-1)(y^4-(b /a)^4)}}\,, 
\end{align}
where the following dimensionless quantities have been introduced, 
\[
y\equiv \frac{r}{\rc}\,, \qquad  a \equiv \frac{\rc}{r_0}\,, \qquad b\equiv \frac{\rt}{r_0}\,. 
\]
By putting (\ref{diff}) into (\ref{L}) and removing the derivative of $r(\sigma)$\,, the classical action is evaluated.  
Then the potential energy (PE) between the W-bosons including the static energy (SE) is obtained as 
\begin{align}
V_{\rm PE+SE}=2\tf\int^{x/2}_0\!\!\dd r\,\mathcal{L} 
=2\tf r_0 a\int_{1}^{1/a} \!\dd y \, \frac{y^4}{\sqrt{(y^4-1)(y^4-(b/a)^4)}}\,. 
\end{align}

\medskip 

Let us here comment on the $a\rightarrow b$ limit (which corresponds to $x\rightarrow \infty$). 
Then the potential is given by 
\begin{align}
V_{\rm PE+SE}
=\tf  \left( \frac{r_0}{L}\right) ^2b^2x+2\tf r_0b \left( \frac{1}{b}-1\right)\,. 
\end{align}
The first term represents a linear potential and the string tension $\sigma_{\rm st}$ 
(not confusing with the string tension $\tf$) is given by 
\[
\sigma_{\rm st} =\tf \left( \frac{\rt}{L}\right) ^2\,,
\]
and the well known result is reproduced. The second term is rewritten as 
\[
2 \tf (r_0-\rt ) = 2m_{\rm W}\,, 
\] 
where $m_{\rm W}$ is the W-boson mass. Hence it can be understood as the static energy of a pair of W-bosons. 

\medskip

Thus the total potential energy $V_{\rm tot}$\,, including the energy of the external electric field, 
is given by\footnote{For the $b=0$ case, the same potential with a specific value of 
the electric field is discussed in \cite{KL}. } 
\begin{align}
V_{\rm tot} &=V_{\rm PE+SE}-Ex \notag \\
&=2\tf r_0a \int_{1}^{1/a} \!\dd y \, \frac{y^4}{\sqrt{(y^4-1)(y^4-(b/a)^4)}}-
\frac{2\tf \alpha r_0}{a}\int_1^{1/a} \! \! \frac{\dd y}{\sqrt{(y^4-1)(y^4-(b /a)^4)}}
\end{align}
where we have introduced the following quantities, 
\[
\alpha \equiv \frac{E}{E_{\rm c}}\,, \qquad E_{\rm c}\equiv \tf \frac{r_0^2}{L^2}\,.
\] 
Here $E_{\rm c}$ corresponds to the critical electric field obtained from the DBI action. 
Note that $E_{\rm c}$ is not modified even after the compactification, 
because the electric field is turned on the $x^1$-direction while 
the $x^3$-direction is compactified.   

\begin{figure}[tbp]
\begin{center}
\includegraphics[scale=.35]{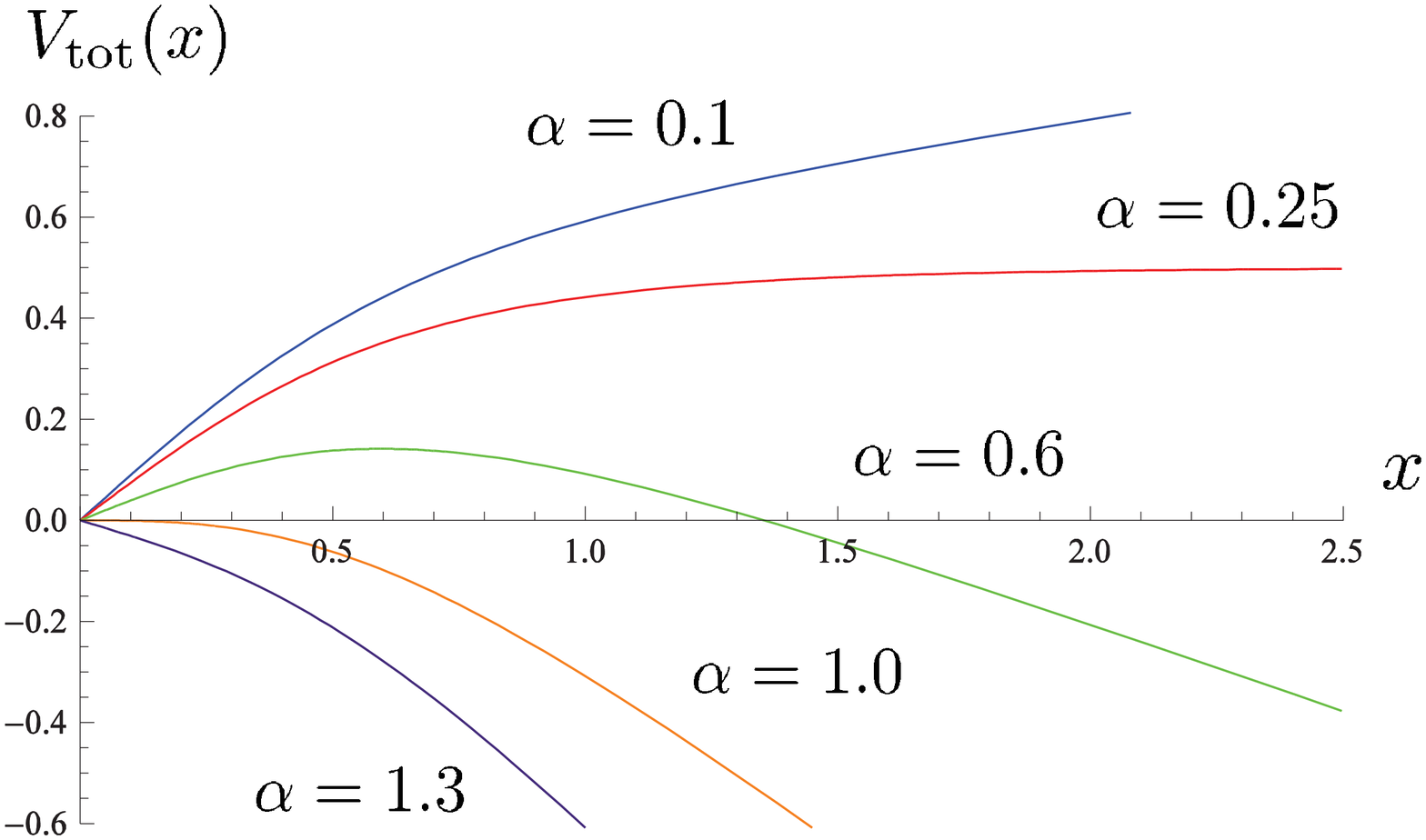}
\vspace*{-0.5cm}
\caption{\footnotesize The numerical plots of the total potential with $b=0.5$ and 
$2L^2/r_0=2\tf r_0=1$\,. 
The blue line is for $\alpha=0.1$\,. There is no zero other than the origin 
and hence the Schwinger effect does not occur. The red line is for $\alpha=0.25$\,. 
This values corresponds to $b^2 =0.25$ and 
the potential becomes flat as $x\to \infty$\,. 
For the values of $\alpha$ between $0.25$ and $1.0$\,, there is the potential barrier 
and the Schwinger effect can occur as tunneling process. 
Just for $\alpha=1.0$ (the orange line), the barrier vanishes 
and the system becomes unstable catastrophically.  
\label{D3-plots:fig}}
\end{center}
\end{figure}

One can see the shape of the total potential numerically. The total potential with $b=0.5$ 
is plotted for $\alpha=0.1$, $0.25$, $0.6$, $1.0$ and $1.3$ in Fig.\,\ref{D3-plots:fig}. 
The potential behavior for the values of $\alpha > 0.25$ is the same as in the Coulomb phase \cite{SY2}. 
The finiteness at the origin is quite similar to the one in non-linear electrodynamics \cite{NL}. 
A remarkable point intrinsic to the confining phase is that the potential becomes flat as $x \to \infty$ 
when $\alpha=0.25$\,. This value corresponds to $b^2=0.25$\,. 
The string tension of confining strings is balanced with the electric field and thus the potential becomes 
flat. Below $\alpha=0.25$\,, 
there is no zero in the potential other than the origin. This means that 
the confining string tension dominates the electric field and no Schwinger effect occurs. 

\subsubsection*{The analytic evaluation of the potential behavior}

So far, we have argued the potential shape numerically, but it is still important to evaluate 
the values of the critical electric fields analytically. 

\medskip 

First of all, let us consider the critical electric field above which the Schwinger effect can occur.  
It is helpful to rewrite the total potential as follows: 
\begin{align}
V_{\rm tot} 
&=\tf  \left( \frac{r_0}{L}\right) ^2(b^2-\alpha)x+\frac{2\tf r_0}{a} \int_{1}^{1/a} \!\dd y \,
 \frac{a^2y^4-b^2}{\sqrt{(y^4-1)(y^4-(b/a)^4)}}\,.  
\end{align}
We would like to show that the potential becomes flat around $x \to \infty$ when  $\alpha = b^2$\,.
Due to the condition $\alpha=b^2$\,, the first term vanishes. 
In order to see the potential behavior around $x\to \infty$\,, 
we take the limit $a \to b$\,. Then the second term becomes constant. After all, the derivative of the total potential 
vanishes. Thus it has been shown analytically that the total potential vanishes around $x\to \infty$ 
when $\alpha = b^2$\,. That is, when the electric field exceed 
\begin{equation}
E_{\rm s}\equiv \tf \left( \frac{r_0}{L}\right)^2b^2 =b^2 E_{\rm c}\,,
\end{equation}
the total potential become flat as $x\to \infty$\,.
Thus $E_{\rm s}$ is regarded as the critical value above which the Schwinger effect is allowed 
to occur as a tunneling process.

\medskip 

Next let us consider the critical electric flux above which the potential barrier vanishes. 
Then it is convenient to rewrite the total potential as follows:
\begin{align}
V_{\rm tot}
&=\tf \left( \frac{r_0}{L}\right) ^2(1-\alpha)x+G(a)\,, \\
&G(a)\equiv 2\tf r_0 \int_{1}^{1/a} \!\dd y \,
 \frac{ay^4-1/a}{\sqrt{(y^4-1)(y^4-(b/a)^4)}}\,.
\end{align}
Here $G(a)$ is a negative-definite and monotonically increasing function. 
By differentiating $V_{\rm tot}$ with respect to $x$\,, we obtain 
\begin{align}
\frac{\dd V_{\rm tot}}{\dd x}
=\tf \left( \frac{r_0}{L}\right) ^2(1-\alpha) +\frac{\dd G}{\dd a}\cdot \frac{\dd a}{\dd x}\,.  
\label{pot-der}
\end{align}
The first term in (\ref{pot-der}) vanishes when $\alpha=1$\,.  
Then the derivatives are given by 
\begin{align}
\frac{\dd G}{\dd a}&=2\tf r_0 \left[ \int_{1}^{1/a} \!\dd y \,
 \left( \frac{y^4+1/a^2}{a^2\sqrt{(y^4-1)(y^4-(b/a)^4)}} - 
\frac{2b^4(ay^4-1/a)}{a^5\sqrt{(y^4-1)(y^4-(b/a)^4)^3}}\right)  \right.\notag \\
&\left. \qquad -\frac{1/a^2-1}{a^3\sqrt{((1/a)^4-1)((1/a)^4-(b/a)^4)^3}}\right]\,, \label{der1} \\
\frac{\dd x}{\dd a}&=-\frac{2L^2}{r_0}
\left[ \int_{1}^{1/a} \!\dd y \,
 \left( \frac{1}{\sqrt{(y^4-1)(y^4-(b/a)^4)}} - 
\frac{2b^4}{a^5\sqrt{(y^4-1)(y^4-(b/a)^4)^3}}\right)  \right.\notag \\
&\left. \qquad +\frac{1}{a^3\sqrt{((1/a)^4-1)((1/a)^4-(b/a)^4)^3}}\right]\,. \label{der2}
\end{align}
Now it is an easy task to check that the second term in (\ref{pot-der}) 
vanishes around $x=0$ (i.e. $a=1$)\,. When $a=1$\,, the integrals in (\ref{der1}) and (\ref{der2}) vanish. 
Non-integral parts in both (\ref{der1}) and (\ref{der2}) diverge as $a \to 1$\,. 
But the divergence is canceled out in the expression (\ref{pot-der})\,. 
Thus the derivative of the potential (\ref{pot-der}) vanishes around $x=0$ 
when $\alpha=1$\,, as we haven seen in the numerical plots. 

\medskip 

It is worth noting the relation to the previous result obtained in \cite{SY2}. 
The case with $b=0$ corresponds to the zero temperature case, and hence 
the above argument supports the value of the critical electric field 
numerically evaluated in \cite{SY2}. For the finite temperature case, 
the value of the electric field numerically shown in \cite{SY2} is supported in the same way.

\section{Confining D4-brane background}

The next is to consider the D4-brane background case. 
The analysis is almost parallel to the D3-brane case in the previous section. 

\medskip 

The metric of the confining D4-brane background is given by 
\begin{align}
\dd s^2&=\left(\frac{r}{L}\right)^{3/2} \left( -(\dd x^0)^2+ \sum _{i=1}^3(\dd x^i )^2+f(r)(\dd x^4)^2 \right)
+\left(\frac{L}{r}\right)^{3/2}\! \! \! f(r)^{-1}\, \dd r^2+L^{3/2}\sqrt{r}\, \dd \Omega _4 ^2\,, \notag \\
f(r)&=1-\frac{\rt^3}{r^3}\,, \qquad L^3=\pi \gs N\alpha^{ \prime 3/2}\,. 
\label{metric2} 
\end{align}
When $\rt =0$\,, the geometry is reduced to the near-horizon geometry of a stack of $N$ D4-branes. 
The parameter $\rt$ plays a role of temperature again. 

\medskip 

\begin{figure}[tbp]
\begin{center}
\includegraphics[scale=.35]{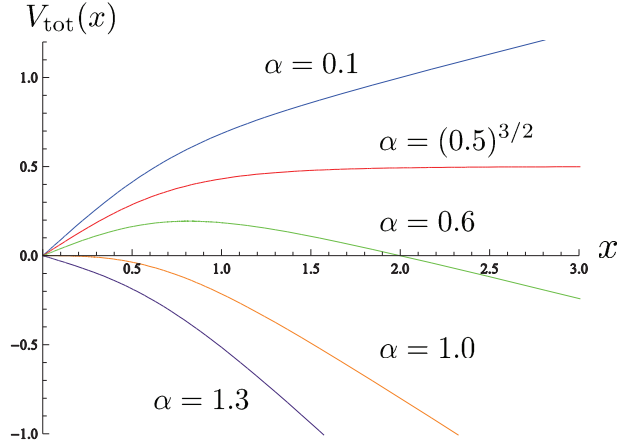}
\vspace*{-0.5cm}
\caption{\footnotesize The numerical plots of the total potential with $b=0.5$ and 
$2L^{3/2}/r_0^{1/2}=2\tf r_0=1$\,. 
The blue line is for $\alpha=0.1$\,. There is no zero other than the origin 
and hence the Schwinger effect does not occur. The red line is for $\alpha=(0.5)^{3/2}$\,. 
This values corresponds to $b^{3/2} =(0.5)^{3/2}$ and 
the potential becomes flat as $x\to \infty$\,. 
For the values of $\alpha$ between $(0.5)^{3/2}$ and $1.0$\,, there is the potential barrier 
and the Schwinger effect can occur as tunneling process. 
Just for $\alpha=1.0$ (the orange line), the barrier vanishes 
and the system becomes unstable catastrophically.  
\label{D4-plots:fig}}
\end{center}
\end{figure}

We are concerned with the Nambu-Goto string action on the background (\ref{metric2}).  
The Lagrangian is given by 
\begin{equation}
{\cal L} =\sqrt{\frac{1}{1-\rt^3/r^3}\left(\frac{\dd r}{\dd \sigma}\right)^2+\frac{r^3}{L^3}}\,. 
\end{equation}
Since $\mathcal{L}$ does not depend on $\sigma$ explicitly, a conserved quantity can be constructed 
as the Hamiltonian with respect to $\sigma$\,, as in the D3-brane case. 
As a result, the following relation is obtained, 
\begin{align}
\frac{r^3}{\displaystyle \sqrt{\frac{1}{1-\rt^3/r^3}\left(\frac{\dd r}{\dd \sigma}\right)^2
+\frac{r^3}{L^3}} }=\rc ^{3/2}L^{3/2}\,, \label{cond-D4}
\end{align}
where we have assumed that the boundary condition at $\sigma =0$\,, 
\begin{equation}
\frac{\dd r}{\dd \sigma} =0\,, \qquad  r=\rc~~~~(\rt <\rc < r_0)\,. \label{bc-D4}
\end{equation}
The condition (\ref{cond-D4}) can be rewritten like the following differential equation, 
\begin{align}
\frac{\dd r}{\dd \sigma}=\frac{1}{L^{3/2}}\sqrt{(r^3-\rt^3)\left(\frac{r^3}{\rc ^3}-1\right)}\,. 
\label{diff-D4}
\end{align}
By solving (\ref{diff-D4}) under the condition (\ref{bc-D4})\,, 
the distance $x$ between W-bosons is given by  
\begin{align}
x
&=\frac{2L^{3/2}}{r_0^{1/2}a^{1/2}}\int_1^{1/a}\! \! \frac{\dd y}{\sqrt{(y^3-1)(y^3-(b /a)^3)}}\,. 
\end{align}
By using (\ref{diff-D4}) and rewriting the Nambu-Goto action, 
the potential energy including the static energy is evaluated as  
\begin{align}
V_{\rm PE+SE}
&=2\tf r_0a \int_{1}^{1/a} \!\dd y \, \frac{y^3}{\sqrt{(y^3-1)(y^3-(b/a)^3)}}\,. 
\end{align}
Then the total potential is given by\footnote{For the $b=0$ case, 
the same potential with a specific value of the electric field is discussed in \cite{KL}. }  
\begin{align}
V_{\rm tot}&=V_{\rm PE+SE}-Ex  \\
&=2\tf r_0a\int_{1}^{1/a} \!\dd y \frac{y^3}{\sqrt{(y^3-1)(y^3-(b/a)^3)}}
-\frac{2\tf r_0\alpha }{a^{1/2}} \int_1^{1/a}\! \! \frac{\dd y}{\sqrt{(y^3-1)(y^3-(b /a)^3)}}\,, \notag
\end{align}
where we have introduced the following quantities
\[
\alpha \equiv \frac{E}{E_{\rm c}}\,,\qquad E_{\rm c} \equiv \tf \left( \frac{r_0}{L}\right)^{3/2} \,.
\]
The shapes of the potential for $\alpha=0.1$, $(0.5)^{3/2}$, $0.6$, $1.0$ and $1.3$ 
are plotted in Fig.\,\ref{D4-plots:fig}. In total, the qualitative behavior of the potential 
is the same as in the case of D3-brane. The potential becomes flat as $x \to \infty$ 
when $\alpha=(0.5)^{3/2}$\,. This value corresponds to $b^{3/2}=(0.5)^{3/2}$\,. 
The string tension of confining strings is balanced with the electric field and 
thus the potential becomes flat. The confining string tension dominates 
the electric field below $\alpha=b^{3/2}$ and again no Schwinger effect occurs.


\subsubsection*{The analytic behavior of the potential}

Let us analyze the potential behavior analytically. The analysis is similar to the D3-brane case. 
The total potential can be rewritten as
\begin{align}
V_{\rm tot}
=\tf \left( \frac{r_0}{L}\right)^{3/2}(b^{3/2}-1)x+2\tf r_0
\int_{1}^{1/a} \!\dd y \frac{ay^3-b^{3/2}/a^{1/2}}{\sqrt{(y^3-1)(y^3-(b/a)^3)}}\,. \notag
\end{align}
One can immediately check that the total potential is flat around $x\to \infty$\,, 
when the electric field is above 
\[
E_{\rm s}=
b^{3/2} E_{\rm c}\,.
\]

\subsubsection*{The region of validity}

Before closing this section, let us discuss the validity region of our computation. 
Note that there is no restriction in the case of D3-brane under the standard 
condition 
\begin{eqnarray}
N\to \infty\,, \qquad \lambda \gg 1\,. \label{sg-cond}
\end{eqnarray} 
For the D4-brane background, apart from the condition (\ref{sg-cond}), 
the radial direction is restricted to some region so that the supergravity description is good. 
The region for the D4-brane case for $b =0$ is shown as \cite{IMSY} 
\begin{eqnarray}
\frac{1}{N\gs}\sqrt{\alpha '} \ll r\ll \frac{N^{1/3}}{\gs}\sqrt{\alpha '}\,.  \label{D4-cond}
\end{eqnarray}
That is, the intermediate region of $r$ is allowed for the supergravity description. 

\medskip 

In the present case we should be careful for the two locations in the radial direction, 
1) the position of the probe D4-brane $r_0$ and 2) the tip of the string world-sheet $\rc$\,. 
First of all, the probe D4-brane is assumed to be put in the region (\ref{D4-cond})\,. 
Then the problem happen when the tip hit on the lower bound of the condition (\ref{D4-cond})\,. 
This implies that the upper bound exists for the range of the distance $x$ between the W-bosons, 
though it seems difficult to evaluate the upper bound analytically. 
This is the scenario for the $b=0$ case. 

\medskip 
 
For $b\neq 0$\,, the lower bound of the condition (\ref{D4-cond}) may be modified. 
When $\rt < \sqrt{\alpha '}/N\gs$\,, there is no modification and 
the previous argument holds. However, when $\rt > \sqrt{\alpha '}/N\gs$\,, 
the lower bound is replaced by $\rt$ and thus the condition for the supergravity approximation 
does not leads to the upper bound of $x$\,.

\section{Conclusion and discussion}

We have performed the potential analysis for confining D3-brane and D4-brane backgrounds. 
For both cases, we have found the critical electric field above which 
the potential barrier vanishes and the system becomes unstable catastrophically. 
An intriguing point is that no Schwinger effect occurs when the electric field is smaller than 
the confining string tension. In other worlds, the tunneling process is allowed 
when the electric field dominates the confining string tension.  

\medskip 

The next interesting problem is to consider the Schwinger effect in QCD-like theories 
with the holographic setup.  
For this purpose, it is necessary to proceed the analysis furthermore.  
An intriguing issue is to argue how to realize non-abelian Schwinger effect \cite{na1,na2,na3}
in the holographic framework, for example, the Sakai-Sugimoto model \cite{SS}. 
We believe that the understanding obtained here would be a key ingredient in this direction.

\section*{Acknowledgments}

We would like to thank H.~Shimada, H.~Suganuma and F.~Sugino for useful discussions. 
This work was also supported in part by the Grant-in-Aid 
for the Global COE Program ``The Next Generation of Physics, Spun 
from Universality and Emergence'' from MEXT, Japan.

\end{document}